\documentclass[11pt]{article}

\usepackage{times}
\usepackage{a4}
\usepackage{verbatim}
\usepackage{graphicx}
\usepackage{amssymb,amsmath}
\usepackage{xspace,hyperref}
\usepackage[affil-it]{authblk}

\usepackage{natbib}
\bibpunct{(}{)}{;}{a}{,}{,}
\newcommand{\ro}[1]{\ensuremath{\textrm{#1}}}
\newcommand{\ten}[1]{\ensuremath{\times 10^{#1}}}

\newcommand{\aT}{\ensuremath{\boldsymbol{\alpha}_T}\xspace}

\title{The Fine-Tuning of the Universe for Life}

\author{Luke A. Barnes%
  \thanks{Electronic address: \texttt{L.Barnes2@westernsydney.edu.au} \\}}
\affil{School of Science \\
Western Sydney University \\
Penrith, NSW 2751, Australia}

\begin{document}
\maketitle

\section{Introduction: Fine-Tuning in Physics}

When a physicist says that a theory is \emph{fine-tuned}, they mean that it must make a suspiciously precise assumption in order to explain a certain observation. This is evidence that the theory is deficient or incomplete. As a simple example, consider a geocentric model of the Solar System. Naively, at any particular time, the Sun and planets could be anywhere in their orbits around the Earth. However, in our night sky, Mercury is never observed to be more than 28$^\circ$ from the Sun, and Venus is never seen more than 47$^\circ$ from the Sun.

\begin{figure*} \centering
		\includegraphics[width=0.7\textwidth]{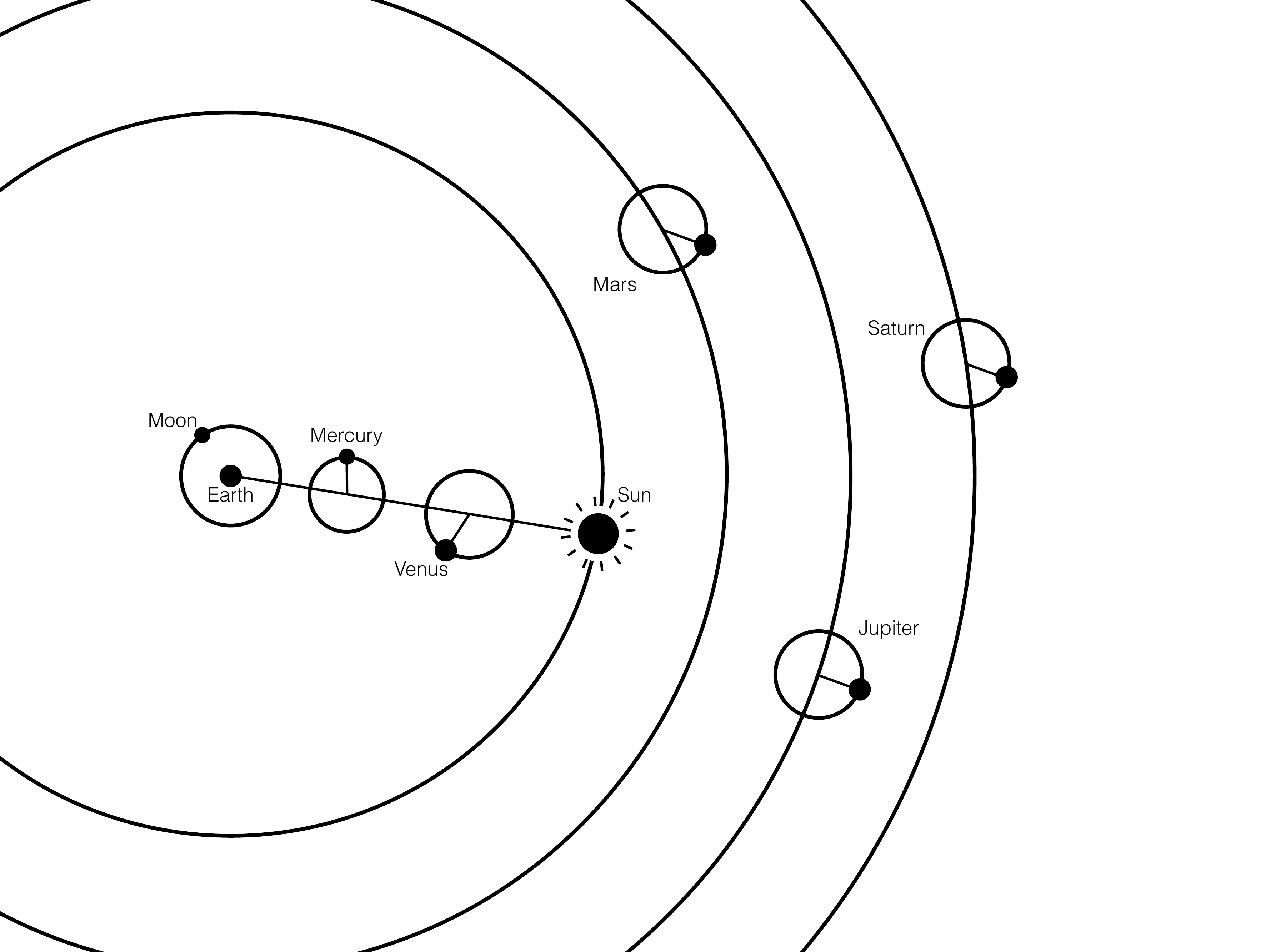}
		\caption{The Ptolemaic model of the Solar System. All the planets travel on epicycles (smaller circles) in orbit around the Earth, but in the case of Mercury and Venus it must be assumed that the centres of their epicycles are fixed to a line that connects in the Earth and the Sun. In this way, Mercury and Venus are never seen too far from the Sun. But in the context of the model, this assumption is unmotivated and suspiciously precise. This presents an opportunity for another model to explain this data more naturally: Mercury and Venus are never seen too far from the Sun because the planets orbit the Sun, not the Earth.}
\label{fig:Ptolemy}
\end{figure*}

Can a geocentric model explain this observation? Yes, but only by adding a postulate. In Ptolemy's geocentric model, Mercury and Venus travel on epicycles, and those epicycles are centred on a line joining the Earth to the Sun (Figure \ref{fig:Ptolemy}). This explains the data, so the model does not fail. However, in the context of the model, this assumption is unmotivated and suspiciously precise. Given only that the planets and Sun orbit the Earth, there is no reason to expect such an arrangement.

This fine-tuning of the geocentric model doesn't necessarily mean that it is wrong, but it should make us wary. We should search for a model in which the data is explained more naturally: Mercury and Venus are never seen too far from the Sun because the planets orbit the Sun, not the Earth.\footnote{This isn't how it happened historically, but it does illustrate the principle.}

Similar arguments play an important role in modern cosmology and particle physics. A standard cosmology textbook case for cosmic inflation goes as follows \citep[e.g..][]{Peacock1998}. In the standard model of cosmology, the geometry of the universe can be negatively curved, flat, or positively curved, depending on whether the density universe is less than, equal to, or greater than the \emph{critical density}. In this model, two facts seem to be in tension with each other. Firstly, the matter in the universe causes the density of the universe to evolve away from critical. Secondly, observations tell us that the density of the universe is very close to critical.

What about in the past? If we extend the model back to nucleosynthesis, about 1 second after the beginning, then the density of the universe must be within one part in $10^{15}$ of the critical density in order to still be close to critical today. The further we push back, the closer the constraint: at the Planck time, it is one part in $10^{55}$. As with Ptolemy's model, the standard model of cosmology can explain the data, but only with an unmotivated and suspiciously precise assumption. We must simply assume that the density of the universe was extremely close to critical in its earliest moments. This motivates inflationary models, in which a early period of accelerating expansion drives \emph{towards} critical density (see Ijjas in this volume).

A second example comes from particle physics \citep{Dine2015}. The observed mass of the Higgs particle can be written in terms of a ``bare'' value and quantum corrections. These quantities are independent in the model. However, the size of the quantum corrections diverges quadratically with the scale up to which the effective theory can be trusted. Dine says, ``if the cutoff is the Planck scale, this correction is enormous \ldots about thirty four orders of magnitude larger than [the observed value], corresponding to a fine tuning of the bare parameters against the radiative correction at the part in $10^{34}$ level.''

Again, the model can explain the observed value, but only by making the unmotivated and suspiciously precise assumption that the bare value almost perfectly cancels out the quantum corrections \citep[see also][]{Donoghue2007}. Particle physicists tend to call this situation ``unnatural'' rather than fine-tuned, but it's a similar idea. As Dine notes, ``naturalness has for many years been a guiding principle in the search for physics beyond the Standard Model''.

The assumptions underlying these arguments have been the subject of much theoretical attention, but the logic is quite widely accepted. The cosmological constant problem, the flatness problem, the big- and little- hierarchy problems of particle physics (see Jacquart in this volume) and the strong CP problem (see Ijjas in this volume) can be framed as fine-tuning problems.

One particular case of fine-tuning is particularly striking. The data in question are not the precise measurements of cosmology or particle physics, but a more general feature of our universe: it supports the existence of life. Before we look at this in more detail, it will be helpful to place fine-tuning in the context of Bayesian approaches to testing physical theories.

\section{Bayesian Accounts of Fine-Tuning}

The Bayesian approach to probability theory views probabilities as quantifying the \emph{degree of plausibility} of some proposition, given other propositions. Bayesians have argued that the familiar probability axioms of \citet{Kolmogorov1933} (or similar) also apply to degrees of plausibility. This can be shown via Dutch book arguments, representational theorems that trace back to \citet{Ramsey1926}, or (more common among physicists) the theorem of \citet{Cox1946}, which proposes that degrees of plausibility obey some intuitive desiderata \citep[see also][]{Jaynes2003,Caticha2009,Knuth2012}.

In the Bayesian approach, physical theories are tested as follows. Let,
\begin{itemize}  \setlength{\itemsep}{-2pt}
\item $T$ = the proposed theory to be tested. As a concrete example, $T$ may represent a set of symmetry principles, from which we can derive the mathematical form of a Lagrangian (or, equivalently, the dynamical equations), but not the values of its free parameters.
\item $U$ = our observations of this Universe.
\item $B$ = everything else we know. For example, we treat the findings of mathematics and theoretical physics as given, so these are included in $B$. As I have defined it for our purposes here, the information in $B$ does not give us any information about which possible world is actual. The theoretical physicist can explore models of the universe mathematically, without concern for whether they describe reality.
\end{itemize}
We then would like to know how plausible $T$ is, in light of everything that we know $UB$. If the \emph{posterior} probability $p(T|UB)$  --- read ``the probability of $T$ given $U$ and $B$" --- were to descend to us on a cloud from the heavens, then our job would be done. Alternatively, we may need some help in calculating the posterior, and so we turn to Bayes Theorem,
\begin{equation} \label{eq:bayesTUB}
p(T|UB) = \frac{p(U|TB) ~ p(T | B)} {p(U | B)} ~.
\end{equation}

If the theory in question has free parameters, which we generically denote $\aT$, then we must take into account our lack of knowledge of these parameters in evaluating the \emph{likelihood} of the data given the theory $p(U|TB)$. We can think of this as dividing the theory into a large number of sub-theories, each with a different value of the free parameters. To calculate the likelihood, we need to average over these sub-theories --- this is known as marginalizing over nuisance parameters. Sub-theories that can account for the data bring the average up, and sub-theories that can't bring the average down.

As a simplistic model, suppose a free parameter varies uniformly over a range $R$, but only a small range $\Delta \aT$ is consistent with the data. Then the theory's likelihood is penalized by a factor $\Delta \aT / R$. The smaller the range of free parameters that accounts for the data, relative to the range of the parameters dictated by the theory, the more the likelihood is penalized. Fine-tuning can be translated directly into improbability within a Bayesian approach \citep[see also][and references therein]{Aguirre2007,Fowlie2014,Barnes2017}.

\section{The Fine-Tuning of the Universe for Life}

Part of exploring any physical model is calculating the effect of varying its free parameters. As we have seen, this is necessary for calculating the likelihood of the data given the theory (via marginalizing), and so this can tell us whether the theory is fine-tuned or not. Beginning in the 1970's, physicists noted that seemingly small changes to the fundamental constants of nature and the initial conditions of the cosmos not only brought our models in conflict with precise measurements; they described universes in which no life form could exist. The complexity and stability required by any known or thus-far conceived form of life could be rather easily erased.

This \emph{fine-tuning of the universe for life} was first investigated by \citet{Carter74}, \citet{Silk1977}, \citet{Carr79}, \citet{Davies1983}, and \citet{BT86}, and has been reviewed recently by \citet{Hogan2000}, \citet{Barnes2012}, \citet{Schellekens2013} and \citet{Lewis2016}. We will consider a few examples.

\subsection{The Cosmological Constant}

The cosmological constant problem is described in the textbook of \citet{BM06} as ``arguably the most severe theoretical problem in high-energy physics today, as measured by both the difference between observations and theoretical predictions, and by the lack of convincing theoretical ideas which address it.'' The problem is as follows. Quantum field theory describes particles as configurations of a field. There is a particular configuration of the field that corresponds to a state with zero particles; this is known as the vacuum state. Because the field is still there, we can ask: how much energy is contained in the vacuum?

The absolute energy of the field doesn't effect the interactions of the standard model of particle physics, which depend only on energy differences. But gravity, on Einstein's theory, responds to the absolute amount of energy. In a homogeneous and isotropic universe, vacuum energy has the same effect as Einstein's cosmological constant. When cosmologists speak of the cosmological constant, they usually mean the sum of the ``bare'' cosmological constant in Einstein's equation and all the forms of energy in the universe that behave in the same way. This is the quantity that is constrained by cosmological data. In Planck units ($\hbar = G = c = 1$) and expressed as a density, the observed cosmological constant has the value $\rho_{\Lambda} \approx 1.2 \ten{-123}$.

We can estimate the contribution to the energy in the vacuum from a given quantum field. Loosely speaking, even in the vacuum state, virtual particles will be created and annihilate, forming loops in a Feynman diagram. The vacuum energy depends on the energy scale up to which we trust the theory to describe this process. Even if we only consider well-understood fields (e.g. the electron field) up to energy scales that have been thoroughly investigated by experiment (say, $\sim 100$ GeV), the contribution to the vacuum energy is $\sim 10^{-68}$, or 55 orders of magnitude larger than the observed value. If we extend the range of our theory up to a popular energy scale where new physics is expected, the supersymmetry scale, then the contribution to the vacuum energy is $\sim 10^{-64}$. If we extend all the way to the Planck scale, where cannot trust our theories because they do not account for quantum gravity effects, the contribution to the vacuum energy is $\sim 1$, 123 orders of magnitude larger than the observed value.

This is a fine-tuning problem. Quantum field theory and general relativity \emph{can} explain the small observed value of the cosmological constant, but only by supposing that the different (positive and negative) contributions to the vacuum energy from each quantum field happen to cancel each other to 123 decimal places. This requires an unmotivated but suspiciously precise coincidence between a number of independent factors.

As an example of fine-tuning \emph{for life}, the cosmological constant problem is a near-perfect storm.
\begin{itemize}
\item It's actually several problems. Each quantum field –-- electron, quark, photon, neutrino, etc. –-- adds a very large (positive or negative) contribution to the vacuum energy of the universe.
\item General Relativity won't help. Einstein's theory links energy and momentum to spacetime geometry. It does not dictate what energy and momentum exists in the universe. Universes that are no good for life are perfectly fine by the principles of General Relativity.
\item Particle physics probably won't help. All particle physics processes, being described by quantum field theory, depend only on energy differences; only gravity  responds to absolute energies. Thus, particle physics is largely blind to its effect on cosmology, and thereby life.
\item It isn't just a problem at the Planck scale, so quantum gravity won't necessarily help. As noted above, we don't need to trust quantum field theory all the way up to the Planck energy in order to see the cosmological constant problem. It is entrenched firmly within well-understood, well-tested physics.
\item Alternative forms of dark energy have very similar problems. They usually posit some other kind of field, and so the problem of the vacuum energy of the field remains, unchanged and unsolved. See Jacquart (this volume) for more discussion of dark energy.
\item We can't aim for zero. Before the accelerated expansion of the universe was discovered 1998, it was thought that some principle or symmetry would set the cosmological constant to zero. Even this was a speculative hope, and it has since evaporated.
\item The quantum vacuum has observable consequences, and so cannot be dismissed as mere fiction. In particular, an electron in an atom feels the influence of the quantum vacuum (the \emph{Lamb shift}). Our theory works beautifully for electrons and atoms. Why doesn't cosmic expansion feel the influence of the quantum vacuum?
\item The cosmological constant has a very obvious and definitive effect on the necessary conditions for life. A positive cosmological constant causes the expansion of the universe to accelerate, freezing structure formation. Make the cosmological constant  a few orders of magnitude larger and structure formation freezes before \emph{anything} has formed. The universe will be a thin, uniform hydrogen and helium soup, a diffuse gas where the occasional particle collision is all that ever happens. A very simple way to make a universe lifeless is to make it devoid of any structure whatsoever. Alternatively, a negative cosmological constant causes the universe to recollapse. If the cosmological constant were $-10^{-68}$, then the universe would recollapse $\sim \sqrt{10^{68}} ~ t_\ro{Planck} \sim 10^{-10}$ seconds after the big bang.
\end{itemize}

\subsection{The Parameters of the Standard Model}

The standard model of particle physics has 25 free parameters which are constrained by experiment. Many of these play a crucial role in providing the complexity required by life.

The Higgs field ``gives mass'' to the fundamental particles of the standard model. We can write their masses in terms of the vacuum expectation value (vev) of the field ($v$) as $m_f = \Gamma_f v / \sqrt{2}$, where $\Gamma_f $ is the particle's dimensionless Yukawa parameter. As with vacuum energy, quantum corrections to the bare Higgs vev are predicted to be of the same order as the scale up to which we trust the theory. The observed value of $v = 1.0 \ten{-17}$ is unnaturally small.

Similarly small changes to $v$ significantly affect how particles interact and bind. \citet{DD2008} refine the approach of \citet{Agrawal1998} by considering nuclear binding, and conclude that unless $0.78 \ten{-17} \lesssim v / m_\ro{Planck} \lesssim 3.3 \ten{-17}$ hydrogen is unstable to the reaction $p + e \rightarrow n + \nu_e$ (if $v$ is too small) or else there is no nuclear binding at all (if $v$ is too large).

Similarly, the strengths of the fundamental forces are subject to anthropic constraints. For example, unless $\alpha \lesssim (m_d - m_u) / $141 MeV, the electromagnetic contribution to the mass of the proton causes it to be heavier than the neutron, making the proton unstable \citep{Hogan2000,HN2008}. If the strong force were a few percent weaker, the deuteron would be unbound \citep{Pochet1991}. The first step in stellar burning would require a three-body reaction to form helium-3. This requires such extreme temperatures and densities that stable stars cannot form: anything big enough to burn is too big to be stable \citep{Barnes2017a}\footnote{The fine-tuning required for stable, life-powering stars has been clarified by recent work by \citet{Adams2008,Barnes2015,Adams2016,AG16,AG17}.}. Weaken the strong force by a few more percent, or increase the strength of electromagnetism, and carbon and all larger elements are unstable \citep{BT86}. The parameters of the standard model must walk a tight-rope in order to form stable nuclei and support stable stars.

\subsection{The Dimensionality of Spacetime}

Spacetime is the arena in which physics takes place. At the length scales relevant to nuclei, atoms, stars, and the observable universe, spacetime is described by three dimensions of space and one of time. It is often straightforward to write down our familiar laws of nature in any number of dimensions. For example, in $m$ time dimensions ($t_i$) and $n$ space dimensions ($x_j$), the wave equation is,
\begin{equation}
\sum\limits_{i = 1}^{m} \frac{\partial^2 \rho}{\partial t_i^2} = c_s^2 \sum\limits_{j = 1}^{n} \frac{\partial^2 \rho}{\partial x_j^2} ~,
\end{equation}
for the scalar wave variable $\rho$, and wave speed $c_s$.

Given that we can theoretically explore such universes, what would they be like? This question has been addressed by \citet{ehrenfest1917}, \citet{whitrow1955}, \citet{BT86}, and \citet{Tegmark1997}. It has been known for some time that Newtonian gravity only predicts stable planetary orbits in three space dimensions (Bertrand's theorem). With four space dimensions, for example, slightly non-circular orbits are spiralled, not elliptical --- they would send the planet into the star or off into empty space. The same applies to atomic orbits described by the Schrodinger equation --- there is no stable ground state.

We can also vary the number of time dimensions. In such a universe, an observer will have their own clock that measures time along their worldline; but what would they experience? \citet{Tegmark1997} notes that linear partial differential equations, of which the wave equation is one example and by which many known laws can be approximated locally, have interesting properties when there is more than one time dimension. In our universe, we can approximately predict the behavior of a physical system into the future on the basis of knowledge of our immediate environment. (I don't necessarily mean \emph{predict} in a mathematical sense. A bird ``predicts'' the path of a flying insect to catch it.) But if there were more (or less) than one time dimension, then the problem would be mathematically ill-posed, being infinitely sensitive to the initial conditions. The behaviour of one's environment could not be predicted using only local, finite accuracy data, making storing and processing information impossible.

\section{The Multiverse}

Fine-tuning in physics serves as impetus to search for a better theory, one which can account for the facts in a more natural way, without unmotivated assumptions. But what could naturally explain a \emph{life-permitting} universe?

Perhaps we won the cosmic lottery: a life-permitting universe exists, despite the seemingly overwhelming odds, because the universe as a whole consists of a vast, variegated ensemble of sub-universes --- a \emph{multiverse}.

A viable multiverse model needs a few ingredients. The first is a physical theory that goes beyond the standard models by promoting the constants of nature and initial conditions to dynamic variables. We have some hints about how to do this. The strengths of the fundamental forces of particle physics are a function of energy, and seem to converge at an energy far above our current experiments. This has led to the development of Grand Unified Theories (GUT), in which the strong nuclear force, weak nuclear force, and electromagnetism are manifestations of a single, unified force \citep[see][]{Raby2010}. At low energy, the greater symmetry of the unified field is \emph{spontaneously broken}: the strengths of the forces are not written in stone in the fundamental equations, but rather are a frozen accident. 

There are other ways to promote the constants to variables. In string theory, there is a \emph{landscape} of solutions to the fundamental equations, with the familiar ``constants'' of physics written into the various folds and holes of the extra, \emph{compactified} spatial dimensions \citep{Schellekens2013}. They become free parameters of the \emph{solution} to the equations, rather than appearing in the equations themselves.

The second ingredient of a multiverse theory is a cosmological mechanism to create domains of the universe with different values of the ``constants''. The leading contender today is cosmic inflation: in its earliest moments, the universe expanded at an accelerating rate, driving it towards critical density and laying down the seeds of cosmic structure.

The successful predictions of inflation require only that our observable universe inflated, but it has been argued that inflation will naturally produce a multiverse \citep[see][]{Linde2015}. Most inflationary models posit a form of energy called an \emph{inflaton field} that drives the expansion of the universe. The physics of a quantum field is codified in its \emph{potential}: the dynamics is analogous to a ball rolling on a hill, and the shape of the hill tells us how the motion of the ball depends on the value of the field. For an inflaton field to cause accelerating expansion, it must be rolling slowly on a very flat section of the potential. Inflation ends when the field rolls off the flat section, usually into a valley. As the field oscillates around the bottom of the valley, reheating begins: the energy in the inflaton field is transferred into ordinary matter and radiation, beginning the hot big bang phase.

But the field is a quantum field, and so will not evolve deterministically (depending on your interpretation of quantum mechanics). Somewhat simplistically, consider an inflating region of the universe, in which the inflaton field dominates the energy of the universe and is slowly rolling. While in most of the region the field will roll into the valley and inflation will end, there is a finite probability that the field in some sub-region will evolve to a state further up the slope. This part of the universe will inflate for longer. Because this sub-region keeps growing in size, it will soon be larger than the original region, and so inflation will always continue somewhere. Given a sufficiently large initial inflating region, post-big-bang pockets form in an inflating background.

If the energy scale of reheating is above the symmetry breaking scale of the fields in the universe, then the symmetry will break differently in different sub-universes. This creates a population of sub-universes with different `constants' and big bang `initial' conditions.

The final ingredient is a selection effect \citep{Wall2003,Bostrom02,2006math......8592N}. Consider the prediction of cosmic microwave background (CMB) anisotropies in the standard big bang model, from which cosmologists infer the values of various cosmic parameters. Like any thermodynamic system, there are fluctuations in the recombining plasma. So, in a sufficiently large universe, the probability of \emph{someone} observing the CMB that we see approaches one regardless of the values of the cosmic parameters. If we tested physical models by calculating the probability that \emph{some} observation in the universe matches our actual observations, then any values of the cosmic parameters would do in an infinite universe. We couldn't infer their values from observations. A multiverse would make this problem even worse.

To resolve this problem, remember that we don't just know that \emph{some} observation has taken place, but that a \emph{particular} observer has made an observation. Even if \emph{some} observer sees a misleading CMB, the vast majority won't, justifying our inference. We apply this to the multiverse: that \emph{some} region of the universe permits life is a good start but not sufficient. What will a \emph{typical} observer see? 

The anthropic prediction by \citet{Weinberg1987} of the cosmological constant provides an excellent test case. Given a large enough variety of sub-universes with different values of the cosmological constant, somewhere will have a value that permits structure to form. In such an ensemble, asks Weinberg, what cosmological constant would a typical observer see? There is nothing in fundamental physics as we know it that singles out $\rho_\Lambda = 0$ as a privileged value, so we assume for the moment that for values of $\rho_\Lambda$ much smaller than the `natural' Planck scale, the multiverse produces a roughly uniform distribution of values. Then (considering positive values of $\rho_\Lambda$ for now) what is the largest value of $\rho_\Lambda$ that permits the formation of structure? Weinberg's analytic calculation gives an upper limit of $\rho_{\Lambda,\ro{max}} \approx 550 \rho_0 \approx 3 \ten{-121}$, where $\rho _0$ is the present cosmic mass density. Weinberg made this prediction before observation showed that $\rho_{\Lambda} \approx 1.2 \ten{-123}$.

A typical observer would expect to observe a vacuum energy roughly comparable with the anthropic upper bound. It can't be larger, of course --- there are no observers in those sub-universes to make an observation. Weinberg's calculation gives the upper bound as being two orders of magnitude above the actual value, which is close enough to take the calculation seriously. My colleagues and I are currently repeating Weinberg's calculation with more sophisticated supercomputer models of galaxy formation.

If we had observed a value that was ten orders of magnitude smaller than the upper bound, then we would conclude that one of the assumptions in our model is probably wrong. We would look for a dynamical or symmetry-based explanation, rather than an anthropic one.

This kind of case for the multiverse has been criticized as speculative and untestable. But it should be remembered that such considerations are almost unavoidable in cosmology. Just as the astronomer must understand their telescope before they can understand what they see through it, when a cosmologist models the universe, they are inevitably modeling a system that contains themselves. We cannot pretend to stand outside the universe. Selection effects cannot be ignored. We are not Dr Frankenstein; we are the monster. We have woken up in a laboratory and are trying to understand how it made us.

We can test the multiverse using Bayesian probability theory. In this case, the ``data'' to be explained is the constants of nature. If the fine-tuning for life implies that almost all observers in the multiverse would observe similar constants to what we observe, then this could provide a major advantage for a multiverse hypothesis over theories in which the constants are free parameters \citep{Aguirre2007,Barnes2017}.

One way in which a multiverse theory can fail spectacularly is known as the \emph{Boltzmann Brain problem}. Physical theories predict observations, and so a multiverse model should --- in principle --- be able to predict what \emph{kind} of observer we would expect to be. One striking feature of our status as observers is that we formed through a long, consistently entropy-increasing process: gravitational collapse into galaxies and stars, stellar burning and supernovae, planet formation, and biological evolution. In some multiverses, including Boltzmann's original multiverse \citep{Boltzmann1895}, most observers form via a chance statistical fluctuation. Without a consistent thermodynamic arrow of time, they will not observe records of the processes that formed them \citep{Hartle2004}. They will observe as much free energy around them as is required for their existence as observers, and almost certainly no more.

To be clear, this is not the philosophical ``brain-in-a-vat'' problem: how can I know whether I'm a Boltzmann brain with false memories? This is a more straightforward ``theoretical prediction meets observation'' scenario: a cosmological theory predicts that a typical observer will be a Boltzmann brain, \emph{and} will observe that they are a Boltzmann brain. And that prediction is wrong. Whether multiverse models can naturally avoid this problem is an open question; see, among many others, \citet{Page2006,Linde2007,Banks2007,deSimone2010,Aguirre2011,Nomura2011,Boddy2013,Albrecht2015,Boddy2015}.

Misgivings about the whole multiverse project are hardly surprising. Are the tests of multiverse theories enough to make it scientific? Unobservable sub-universes are very different to unobservable quarks: we can constrain the properties of quarks via experiment, but every other sub-universe in the multiverse could disappear tomorrow and we would never know. The meagre tests of the multiverse ``prove nothing'', say \citet{ES14}, "Fundamentally, the multiverse explanation relies on string theory, which is as yet unverified, and on speculative mechanisms for realizing different physics in different sister universes. It is not, in our opinion, robust, let alone testable.''

A potentially tricky hurdle is the \emph{measure problem}, about which there is an extensive literature. Many multiverse theories imply or assume that there are an infinite number of other sub-universes. Given a finite population, deriving probabilities is straightforward: what fraction of observers see a value of $\rho_\Lambda$ as small as the one we observe? But in an infinite multiverse, we cannot simply count sub-universes.

In particular, once we have a useful definition of an observer, it seems that we should treat them all on equal footing. Think of this as permutation symmetry --- having arbitrarily numbered all the observers (or observer moments), we should be able to shuffle the labels without changing the prediction of the model. But there is no assignment of probabilities to an infinite number of possibilities that respects this symmetry. This is often taken as incentive to assign different probabilities. But it could be argued, and with considerable force, that this means that an infinite multiverse theory cannot justify probabilities and so cannot make predictions. ``In an infinite universe,'' says \citet{Olum2012}, ``everything which can happen will happen an infinite number of times, so what does it mean to say that one thing is more likely than another?''. These are open questions; see, among many others, \citet{1995PhRvD..52.3365V,2006JCAP...01..017G,2007PhRvD..75l3501A,2007JHEP...01..092V,2007JPhA...40.6777V,2008PhRvD..77f3516G,2008JCAP...10..025P,2009PhRvD..79f3513B,deSimone2010,2011CQGra..28t4007F,2012PhRvD..85d5007B,2013JCAP...05..037G,Carroll2017,2017arXiv170800449P}.

\section{After Physics}

In physics, fine-tuning problems afflict theories that seem to be successful, that is, that can account for the data. The problem is not a falsified prediction, as one might expect from a discounted or discarded theory. Recall the lesson of Ptolemy's model. Within the set of possible geocentric planetary systems, an uncomfortably large proportion look very different to our Solar System. A fine-tuning problem is raised by a large set of alternate possibilities. This suggests an interesting thought experiment.

Suppose there is an ultimate theory of physics. At a future International Meeting of Really Important Physicists, Alberta Einstein walks to the chalkboard, scribbles a few equations, and fundamental physics comes to an end. Like chess pieces who had discovered the laws of chess, no deeper rules exist.

By hypothesis, this theory would be consistent with all scientific data. But we may still glimpse a large set of alternate possibilities, and so a kind of fine-tuning problem remains. Even if it contains no free parameters, Alberta's chalkboard will show one particular mathematical equation or structure. We will be faced by a very old question: why \emph{this} universe? Of all the ways the world could have been, why this way? Of all the mathematically consistent chalkboards of equations, why Alberta's?

Obviously, the answer is not yet-another chalkboard of equations. Neither is it more observations of this universe. This is not the kind of question that physics can answer, because we can't prove from any set of equations that they describe reality. Theories don't predict their own success. But if not physics, then what? What do we do when fundamental physics is over?

Perhaps we stop asking questions. Maybe reality doesn't have any ultimate reason for why it is the way it is. Explanations of the physical world reach the ultimate laws, and stop. This is the supposition of \emph{naturalism}: the natural world is all there is. For a modern defence, see \citet{Carroll2016}.

Alternatively, \citet{Tegmark1998} has defended the ``the ultimate ensemble theory'', that ``physical existence is equivalent to mathematical existence". The actual world is not chosen from a set of mathematical possibilities; rather, all mathematical possibilities are equally real, and we are self-aware substructures (SASs) within a particular mathematical structure. A metaphysician might worry about the dissolution of the line between \emph{abstract} and \emph{concrete}. The physicist who tries to test Tegmark's idea via its prediction that ``the mathematical structure describing our world is the most generic one that is consistent with our observations" faces a problem: we need a probability distribution over the set of mathematical structures, but a probability distribution is itself a mathematical structure. Tegmark says that probabilities are ``merely subjective'', but our subjective states of mind are mathematical substructures, too.

By contrast, axiarchism \citep{Leslie1989} and theism \citep[e.g.][]{Swinburne2004,Collins2009} argue that beneath the mathematical structure of our universe is a \emph{reason}: our universe is morally valuable, particularly its embodied, free, conscious agents. Just as Tegmark promotes possibilities to reality on mathematical grounds, axiarchism does so on moral grounds: the world exists because it is good. Theism proposes that God exists necessarily in some sense, and the physical world is the result of God's free choice to create a morally valuable world.

For each of these alternatives, the fine-tuning of the universe for life plays an important role. For Tegmark, the complexity required by any SAS explains why we see this universe/mathematical structure, rather than a simpler one. For axiarchism and theism, fine-tuning for life shows how these ideas could have explanatory power. Given the seemingly extraordinarily small proportion of possibilities that permit the existence of embodied moral agents, the axiarchist and theist can understand something of why Alberta's blackboard is the one has gone to all the bother of existing. Further examination of these alternatives takes us beyond the philosophy of physics.

\section*{Acknowledgments}
Supported by a grant from the John Templeton Foundation. This publication was made possible through the support of a grant from the John Templeton Foundation. The opinions expressed in this publication are those of the author and do not necessarily reflect the views of the John Templeton Foundation.

\end{document}